\documentstyle[a4,12pt,epsf]{article}

\newcommand{\wt}{\widetilde}
\newcommand{\ol}{\overline}
\def\tr{\mathop{\rm tr}}

\newcommand{\rap}[2]
{\setbox1=\hbox{#1}%
\setbox2=\hbox to\wd1{\hss #2\hss}%
\mbox{\rlap{\box1}\box2}}

\begin{document}
\begin{titlepage}

\rightline{\tt hep-th/0204200}
\rightline{UT-02-21}

\begin{center}

\vspace{12ex}

{\Large Large angular momentum closed strings\\[1.5ex]
colliding with D-branes}

\vspace{7ex}

{\large Yosuke Imamura}%
\footnote{E-mail: {\tt imamura@hep-th.phys.s.u-tokyo.ac.jp}}

\vspace{5ex}

 {\baselineskip=15pt
 {\it Department of Physics,  Faculty of Science, University of Tokyo \\
  Hongo 7-3-1, Bunkyo-ku, Tokyo 113-0033, Japan} 
 }

\vspace{10ex}

\end{center}
\begin{abstract}
We investigate colliding processes of closed strings with large angular momenta with D-branes.
We give explicit CFT calculations for closed string states with an arbitrary number of bosonic excitations
and no or one fermion excitation.
The results reproduce the correspondence between closed string states
and single trace operators in the boundary gauge theory
recently suggested by Berenstein, Maldacena and Nastase.
\end{abstract}
\end{titlepage}

\newpage

\section{Introduction}
The AdS/CFT correspondence\cite{adscft} relates large $N$ quantum gauge
theories and gravity or string theories in anti-de Sitter (AdS) spacetime.
We can compute various quantities in Large $N$ Yang-Mills theories
by using gravity in AdS backgrounds.
For example, partition functions of a four dimensional ${\cal N}=4$
supersymmetric Yang-Mills theory with a source term $\int J_i{\cal O}_i d^4x$
for gauge invariant operators ${\cal O}_i$
are computed as classical actions of supergravity solutions
in the ${\rm AdS}_5$ background.\cite{GKP,holography}
Each source $J_i$ of operators ${\cal O}_i$ with conformal dimension $d_i$
is taken into account in the supergravity computation
as a boundary condition
$\phi_i\sim J_ir^{4-d_i}$ on the conformal boundary at $r=0$
for a supergravity field $\phi_i$ corresponding to the operator ${\cal O}_i$.
(The conformal metric $ds^2=(dr^2+\eta_{\mu\nu}dx^\mu dx^\nu)/r^2$ is assumed.)
This idea is based on the fact that the near horizon geometry
of a large number of coinciding D3-branes
is ${\rm AdS}_5\times{\rm S}^5$ and gauge invariant operators
in the Yang-Mills theory on the D3-branes couple to bulk fields.
In the weak coupling regime, the D3-branes can be treated as branes
without thickness in the flat background
and the sources for gauge invariant operators ${\cal O}_i$
are supplied by expectation values of bulk fields $\langle\phi_i\rangle$
via brane-bulk coupling $\int{\cal O}_i\phi_i d^4x$.
In strong coupling regime, this is re-interpreted as boundary conditions
mentioned above.
Although a lot of works had been done about this correspondence
between boundary operators and bulk fields,
only restricted types of operators had been discussed.

Recently, Berenstein, Maldacena and Nastase suggested
correspondence between new class of
boundary operators and bulk fields\cite{BMN}.
They defined a charge $J$ associated with a ${\rm U}(1)$ subgroup
of the ${\rm SU}(4)_R$ symmetry of four dimensional ${\cal N}=4$
supersymmetric Yang-Mills theory
and investigated operators with large $J$ and small $d-J$.
The conformal dimension $d$ of an operator is bounded below by the BPS bound $d\geq J$.
In the context of bulk supergravity, $J$ is regarded as the angular momentum along a certain direction
in ${\rm S}^5$.
Therefore, near BPS operators with small $d-J$ and large $J$
correspond to fields
with near light like momentum in ${\rm AdS}_5\times{\rm S}^5$.
In the Penrose limit\cite{Penrose},
which is a natural limit to study near light-like fields,
the background geometry ${\rm AdS}_5\times{\rm S}^5$ reduces to a PP-wave background.
This background has maximal supersymmetry as ${\rm AdS}_5\times{\rm S}^5$.\cite{ppwave,susy2}
Recently, string on this background was exactly solved
by the Green Schwarz formalism in light-cone gauge\cite{Metsaev}.
In Ref\cite{BMN}, it is shown that the string spectrum on the pp-wave background is
exactly reproduced as single trace operators constructed from $\tr({\cal Z}^J)$
by insertion of fields carrying $d-J=1$.

In the weak coupling regime, in which the D3-branes are
treated as branes without thickness in the flat spacetime,
the suggested correspondence between string states and operators
implies that closed strings with large angular momenta
couple to corresponding single trace operators defined on the D3-branes.
The purpose of this work is to check directly by CFT computation
that such couplings between closed strings
and the operators really exist.
By computing disk amplitudes we explicitly show
that the correspondence given in \cite{BMN} is
correctly reproduced as couplings among
large angular momentum closed strings
and open strings on D3-branes.
Although explicit calculation is given
only in the case that
the string excitation is purely bosonic or includes one fermion oscillator,
it is plausible that the correspondence is proved in the same way
in the general case.

Usually, strings on the pp-wave background are studied
by the light-cone GS formalism\cite{lcgs}.
Here, to use the worldsheet CFT tecnique\cite{FMS}, we use the NSR formalism.
In the light-cone formalism,
the interpretation of the light-cone direction in the boundary CFT is not clear.
This is related to a question discussed in several works:
``where does the dual gauge theory live?''\cite{DGR,LOR,KirPio}
To avoid this problem, and clarify the correspondence between string modes
and single trace operators,
we define `spacelike light-cone' in the next section.

\section{Spacelike light-cone}
Let us consider coinciding D3-branes in the flat Minkowski spacetime.
We denote the spacetime coordinates $X^I$ ($I=0,\ldots,9$)
where $X^\mu$ ($\mu=0,1,2,3$) are Neumann directions and
$X^i$ ($i=4,5,6,7$), $Z=(X^8+iX^9)/\sqrt2$ and $Z^\ast$ are
Dirichlet directions.
We define $J$ as the angular momentum on the $Z$-plane.
We will discuss closed strings with large $J$ compared
with the excitation number $L_0+\wt L_0$.

In \cite{BMN}, the operator-string state correspondence is discussed in the
framework of AdS/CFT correspondence.
In this case, we should take the large 't Hooft coupling limit.
The 't Hooft coupling $\eta=Ng_{\rm YM}^2$ determines the radius of
the AdS$_5$ and large $\eta$ justifies the treatment of the AdS$_5$ background
as a solution of supergravity.
In this paper, we use a completely different approach.
We are going to study the correspondence
as interactions between closed strings and D3-branes
in the flat Minkowski background.
Thus, we should assume the 't Hooft coupling $\eta$
(and the string coupling $g_{\rm str}$) is sufficiently small.
Similarly, we do not have to take the large $N$ limit,
which is necessary in the AdS/CFT framework to decouple
uncontrollable effects of the quantum gravity.
We emphasize that we are not going to consider AdS/CFT correspondence.
We will only study the coupling between closed strings and operators
in a weak coupling gauge theory.

When we compute amplitude,
we consider only colliding
processes satisfying the following conditions.
\begin{itemize}
\item Zero momentum condition: Every open string participating in processes has zero momentum.
\item Least number condition: The number of open string vertices is the smallest for a given closed string state.
\end{itemize}

Because we are assuming the small string coupling constant,
contribution of higher genus worldsheets
are suppressed and we can take a disk amplitude as a leading contribution.
If the number of open string vertices inserted on the boundary of the disk
is $n$, the amplitude is proportional to $g_{\rm str}^{n/2}$.
(The factor $g_{\rm str}$ due to one closed string vertex inserted inside
the disk is included.)
Therefore, the contribution of a diagram with the least number of the open string
vertex dominates.
Therefore we may focus on single trace operators
satisfying the least number condition.

We adopt the zero momentum condition to simplify the computation of amplitudes.
Of cause, if we want to consider local operators,
we should represent them as superpositions of infinite
number of Fourier modes.
Such a treatment is necessary if we discuss correlation functions
of local operators.
We leave this problem for future study.

The zero momentum condition drastically reduces the open string states we have to consider.
Because only massless fields can be on-shell at zero momentum,
it is sufficient to consider states in the ${\cal N}=4$ vector multiplet.
The fields and corresponding vertex operators are summarized in Table \ref{n4vector.tbl}.
We have chosen boson and fermion vertices in picture $0$ and in picture $-1/2$,
respectively.
\begin{table}[htb]
\caption{Massless fields belonging to the ${\cal N}=4$ vector multiplet on D3-branes
and corresponding vertex operators.
$\partial_\perp$ and $\partial_\parallel$ define the
derivatives perpendicular or tangential to the worldsheet boundary.}
\label{n4vector.tbl}
\begin{center}
\begin{tabular}{cccccc}
\hline
\hline
fields & $A_\mu$ & $\phi_i$ & ${\cal Z}$ & ${\cal Z}^\ast$ & $\lambda_a$ \\
\hline
vertex operators & $\partial_\parallel X^\mu$ &
$\partial_\perp X^i$ & $\partial_\perp Z^\ast$ &
$\partial_\perp Z$ & $e^{-\phi/2}S^a$ \\
\hline
\end{tabular}
\end{center}
\end{table}

The zero momentum condition also constrains closed string states.
By the momentum conservation, closed strings cannot carry momenta longitudinal to the D3-branes.
This forbids on-shell plane waves except massless states with constant wave functions.
However, we can consider closed strings carrying non-vanishing angular momenta in the following way.
Vertex operators of closed strings generally have the form
\begin{equation}
{\cal O}_c=V\wt Vf(X^I),
\label{general}
\end{equation}
where $V$ and $\wt V$ represent excitations of left and right-moving parts, respectively,
which are the origin of the spin and the mass of the states,
and $f(X^I)$ is an orbital wave function.
First, let us discuss the wave function $f(X^I)$.
For the vertex operator to be primary operator with conformal dimension $(1,1)$,
the wave function should satisfy the Laplace equation:
\begin{equation}
(\partial_i\partial_i+2\partial_z\partial_{z^\ast}-M^2)f(X^i,Z,Z^\ast)=0.
\label{laplace}
\end{equation}
The wave function does not depend on the longitudinal coordinates $X^\mu$
because of the zero momentum condition.
$M$ in the Laplace equation (\ref{laplace}) is a mass of the state determined by the
conformal dimension of the oscillator parts $V$ and $\wt V$.
If we assume a plane wave function $f(X^I)=\exp(ik\cdot X)$,
only zero momentum mode ($f(X^I)=$const) of massless states
can satisfy (\ref{laplace}).
However, if we do not do so,
there exist many non-trivial wave functions.
What we will consider is a wave function $f_J(Z,Z^\ast)$
depending on $Z$ and $Z^\ast$ and satisfying the $J$ eigenstate equation
\begin{equation}
(Z\partial_Z-Z^\ast\partial_{Z^\ast})f_J(Z,Z^\ast)=Jf_J(Z,Z^\ast).
\label{Jeigenstate}
\end{equation}
Although the function $f_J$ grows exponentially at large $|Z|$,
only the leading term in its power series expansion in $Z$ contributes
amplitudes of processes satisfying the least number condition
and the divergence of $f_J$ does not cause any problems.
Therefore, we can use the wave function $f_J(Z,Z^\ast)$
to make the closed string vertex (\ref{general}).
We normalize the wave function such that the coefficient of the leading term is one.
\begin{equation}
f_J(Z,Z^\ast)=\sum_{k=0}^\infty\frac{(M^2Z^\ast)^k}{k!}\frac{Z^{J+k}}{(J+k)!}.
\end{equation}

Next let us determine physical components of the oscillator factors $V$ and $\wt V$.
We discuss only massless states here.
Excited states can be treated in a similar way.
The vertex operator of the massless NS-NS fields in picture $(-1,-1)$ is
\begin{equation}
{\cal O}_c=V_{-1}[\zeta]\wt V_{-1}[\wt\zeta]f_J,
\label{NSNSvertex}
\end{equation}
where the excitation parts $V_{-1}$ and $\wt V_{-1}$
are given by
\begin{equation}
V_{-1}[\zeta_I]
=\zeta_IV_{-1}^I
=e^{-\phi}(\zeta_I\psi^I),\quad
\wt V_{-1}[\wt\zeta_I]
=\wt\zeta_I\wt V_{-1}^I
=e^{-\wt\phi}(\wt\zeta_I\wt\psi^I),
\end{equation}
and the wave function is $f_J=Z^J$.
The physical components of the polarization vectors $\zeta_I$ and $\wt\zeta_I$
are determined as follows.
The BRS invariance of the vertex operator
requires the following OPE to be regular.
\begin{equation}
j_{\rm BRS}(w)\cdot cV_{-1}f_J(z)
\sim\frac{1}{w-z}\eta c\zeta^I\frac{\partial f_J}{\partial X^I}.
\end{equation}
Because $f_J=Z^J$,
This implies that $\zeta^z=0$.
The gauge transformation of this vertex operator is
\begin{equation}
j_{\rm BRS}(w)\cdot c\Lambda_{-1}(z)
\sim\frac{1}{w-z}ce^{-\phi}\psi^I\frac{\partial\lambda}{\partial X^I},
\label{gaugetr}
\end{equation}
where $\lambda$ is the gauge transformation parameter
depending on the coordinates $X^I$ and $\Lambda_{-1}$ is
the following operator in picture $-1$.
\begin{equation}
\Lambda_{-1}\sim e^{-2\phi}\partial\xi\lambda(X^I).
\end{equation}
If we take $\lambda=Z^{J+1}$, the gauge transformation (\ref{gaugetr}) implies that $\zeta^{z^\ast}$ is
a gauge degree of freedom.
Thus, we conclude that physical components of the vector $\zeta^I$ are
$\zeta^\mu$ ($\mu=0,1,2,3$) and $\zeta^i$ ($i=4,5,6,7$).
In other word, physical degrees of freedom are $J=0$ components of the polarization vector.
The same condition is obtained for the right moving part polarization vector $\wt\zeta^I$
by using $\wt j_{\rm BRS}$.
This situation is quite similar to the light-cone formalism,
in which the wave function include a plane wave factor $e^{ik^-X^+}$ and
physical mode oscillations are transverse to the light-cone coordinates $X^\pm$.
In this sense, we refer to the $Z$-plane as `(spacelike) light-cone' directions.

Similar arguments can be applied to fermionic vertices.
The excitation factor representing the R-vacuum in picture $-1/2$ is
\begin{equation}
V_{-1/2}[u_a]
=u_aV^a_{-1/2}
=e^{-\phi/2}(u_aS^a),
\label{Rvertex}
\end{equation}
where $u_a$ is a $16$-component polarization spinor.
The OPE of the vertex and the BRS current is
\begin{equation}
j_{\rm BRS}(w)\cdot cV_{-1/2}f(z)\sim\frac{1}{w-z}\eta c(u\Gamma^I\ol S)\frac{\partial f_J}{\partial X^I}.
\end{equation}
If we take the massless wave function $f_J=Z^J$,
the BRS invariance condition demands that the spinor $u$ satisfies
\begin{equation}
u\Gamma^z=0.
\end{equation}
This implies that eight components of $u_a$ carrying $J=-1/2$ are physical.

\section{Bosonic excitations}
In the last section, we have determined the massless physical states of a closed string with `light-cone' angular momentum.
They consist of 128 bosonic and 128 fermionic massless physical states.
Similarly to the ordinary light-cone GS formalism,
the whole Fock space of closed string physical states
is constructed by exciting these $256$ vacuum states
by oscillators of eight boson fields carrying $J=0$
and eight fermions carrying $J=1/2$.
In this section, we discuss closed string states excited by only
bosonic oscillators $X_{-n}^\mu$ ($\mu=0,1,2,3$) and $X_{-n}^i$ ($i=4,5,6,7$).
Because we can construct all the $256$ ground states
from one of them as states `excited' by fermionic zero modes,
we here choose one ground state as a `vacuum' state.
It will be specified in the next section.
We represent the vertex operator of the `vacuum' state of a closed string by
\begin{equation}
{\cal O}_c={\cal F}f_J(Z,Z^\ast),
\label{simplest}
\end{equation}
where $\cal F$ is an operator in picture $-2$ consisting of matter fermions and superconfomal ghosts
as will be shown in the next section.

\begin{figure}[htb]
\centerline{\epsfbox{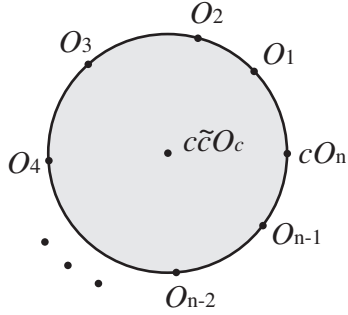}}
\caption{A worldsheet of a closed string colliding with D-branes
and decaying into a large number of open strings}
\label{disk.eps}
\end{figure}

Let us consider a non-vanishing disk amplitude including one closed string vertex (\ref{simplest})
and the least number of open string vertices.
Because the closed string vertex has the factor $Z^J$,
we need at least $J$ open string vertices $\partial_\perp Z^\ast$.
As we see below, this amplitude does not vanish
if ${\cal F}$ is chosen appropriately.
The amplitude we would like to compute is
\begin{equation}
{\cal A}_0=\int d^{J-1}\theta
\langle c\wt c{\cal O}_c(\mbox{center})\cdot c{\cal O}_J(2\pi)\cdot
{\cal O}_{J-1}(\theta_{J-1})\cdots
{\cal O}_2(\theta_2)\cdot{\cal O}_1(\theta_1)
\rangle,
\label{diskamp1}
\end{equation}
where ${\cal O}_i(\theta_i)=\partial_\perp Z^\ast(\theta_i)$. (Figure \ref{disk.eps})
The integral $\int d^{J-1}\theta$ is defined by
\begin{equation}
\int d^{J-1}\theta
\equiv\int^{2\pi}_0d\theta_{J-1}
\int^{\theta_{J-1}}_0d\theta_{J-2}
\cdots
\int^{\theta_2}_0d\theta_1.
\end{equation}
We fix the residual ${\rm SL}(2,{\bf R})$ diffeomorphysm
by putting the closed string vertex
at the center of the disk and the open string vertex ${\cal O}_J$ at the point $\theta=2\pi$
on the boundary.
The amplitude factorizes as
\begin{equation}
{\cal A}_0=
\langle c\wt c(\mbox{center})\cdot c(\theta=0)\rangle
\langle{\cal F}\rangle
\frac{J!}{J}\int d^{J-1}\theta\prod_{i=1}^J\langle Z(\mbox{center})\cdot\partial_\perp Z^\ast(\theta_i)\rangle,
\label{amp0}
\end{equation}
where $J!$ represents the number of combinations of contractions between
$J$ $Z$s and $J$ $\partial_\perp Z^\ast$
and the $1/J$ factor reflects the ambiguity of the choice of
the position-fixed vertex ${\cal O}_J$.
Because the conformal ghost factor $\langle c\wt c\cdot c\rangle$
and the matter fermion and superconformal ghost factor
$\langle{\cal F}\rangle$ give just constant factors,
let us focus on the contribution from contractions of
$Z$ and $Z^\ast$.
The Green function of a scalar field
on the unit disk with Dirichlet boundary condition is
\begin{eqnarray}
\langle\phi(w)\phi(z)\rangle=-\log|w-z|+\log|1-wz^\ast|.
\label{corr}
\end{eqnarray}
From this Green function, we easily obtain
\begin{equation}
\langle\ Z(0)\ \partial_\perp Z^\ast(e^{i\theta})\ \rangle=1.
\label{ZZast}
\end{equation}
By substituting (\ref{ZZast}) into (\ref{amp0}) and carrying out the integrations,
we obtain
\begin{equation}
{\cal A}_0=
\langle c\wt c\cdot c\rangle\langle{\cal F}\rangle(J-1)!I_{J-1}(2\pi).
\label{Amp0f}
\end{equation}
The function $I_n(x)$ is defined as an integration over $n$
ordered variables in the interval $[0,x]$.
\begin{equation}
I_n(x)=
\int_0^xdx_n
\int_0^{x_n}dx_{n-1}
\cdots
\int_0^{x_2}dx_1\cdot1
=\frac{x^n}{n!}.
\end{equation}
The amplitude (\ref{Amp0f}) is interpreted as a coupling between the
closed string state (\ref{simplest}) and the single trace operator $\tr({\cal Z}^J)$.

Let us consider excitation modes of a closed string.
We first discuss excitation by oscillators of Dirichlet directions $X_{-n}^i$ and $\wt X_{-n}^i$ ($i=4,5,6,7$).
The excitation by oscillators of Neumann directions $X_{-n}^\mu$ and $\wt X_{-m}^\mu$ ($\mu=0,1,2,3$) will be
mentioned later.
An excitation by $X_{-n}^i$ ($\wt X_{-n}^i$)
is realized by adding a factor $\partial^nX^i$ ($\wt\partial^nX^i$)
to the massless vertex operator.
Here we discuss only the case with one left-moving excitation $X_{-m}^i$
and one right-moving excitation $\wt X_{-n}^j$.
Generalization to the case with arbitrary number of excitations
is straightforward.
The vertex operator is
\begin{equation}
{\cal O}_c=\frac{1}{m!}\partial^mX^i\cdot
         \frac{1}{n!}\wt\partial^nX^j\cdot
         {\cal F}\cdot Z^J.
\label{11excitation}
\end{equation}
In general, vertex operators obtained in this way are not BRS invariant
and we need to add extra terms in order to make them  BRS invariant.
For example, the massive wave function $f_J$ is not simply $Z^J$
but includes infinite terms
of the form $Z^{J+k}Z^{\ast k}$ (k=1,2,\ldots).
However, these terms do not contribute amplitudes we will consider below
because these extra terms consist of larger number of fields than the
leading term $Z^J$ and some of them do not have partner operators
to be contracted with in amplitudes satisfying the least number condition.
We also need to modify the excitation part
$V\wt V\sim\partial^mX^i\wt\partial^nX^j{\cal F}$ because
operators of the form $\partial^n X^I$, $\wt\partial^nX^I$ and their products
are in general not primary operators.
We have to add extra terms to the vertex operators.
In the computation of amplitudes given below, we neglect such extra terms.
Although we can explicitly show for some simple examples that they indeed
do not contribute the amplitude,
in order to prove the irrelevance of these terms for arbitrary cases,
we need detailed investigation of structure of vertex operators
of circular wave closed strings.
Unfortunately, we have not obtained a strict proof at this point and
we just suppose that these terms do not affect our analysis of
the operator-string state correspondence.

\begin{figure}[htb]
\centerline{\epsfbox{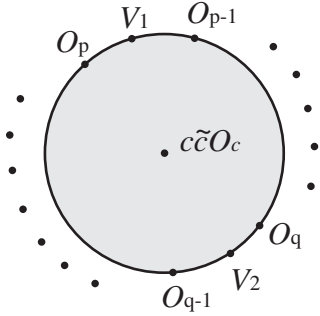}}
\caption{A worldsheet with two extra insertions of vertices
$V_1=\partial_\perp X^i$ and $V_2=\partial_\perp X^j$.}
\label{insert.eps}
\end{figure}

If we replace ${\cal O}_c$ in (\ref{amp0}) by the vertex (\ref{11excitation}),
the amplitude vanishes.
To obtain non-vanishing amplitude,
we need two vertices $\partial_\perp X^i(\varphi)$ and $\partial_\perp X^j(\varphi')$ on the boundary
in addition to the $J$ vertices $\partial_\perp Z^\ast(\theta_i)$.
We insert these two extra vertices into the $p$-th and $q$-th intervals.
(Figure \ref{insert.eps})
Namely, the angles $\varphi$ and $\varphi'$ satisfy
\begin{equation}
\theta_{p+1}>\varphi>\theta_p,\quad
\theta_{q+1}>\varphi'>\theta_q.
\end{equation}
Contribution from the contractions between these extra open string vertices and
excitation factors in the closed string vertex (\ref{11excitation}) is
obtained from the correlation function (\ref{corr}).
\begin{equation}
\frac{1}{m!}\langle\ \partial^mX^i(0)\ \partial_\perp X^{i'}(e^{i\varphi})\ \rangle
=\delta^{ii'}e^{-im\varphi},\quad
\frac{1}{n!}\langle\ \wt\partial^nX^j(0)\ \partial_\perp X^{j'}(e^{i\varphi'})\ \rangle
=\delta^{jj'}e^{in\varphi'}.
\end{equation}
The amplitude is given by
\begin{equation}
{\cal A}=
\langle c\wt c\cdot c\rangle
\langle{\cal F}\rangle
\frac{J!}{J}
\int d^{J-1}\theta
\int_{\theta_p}^{\theta_{p+1}}d\varphi
\int_{\theta_q}^{\theta_{q+1}}d\varphi'
e^{-im\varphi}e^{in\varphi'}.
\end{equation}
(If the directions of the polarizations of the two excitations coincide,
there are one more term coming from the other contraction of boson fields. We here assume $i\neq j$.)
When the number $J$ is very large and both $|p-q|$ and $J-|p-q|$
are the same order of magnitude as $J$, this integral can be estimated
with the help of the dilute gas approximation.\cite{BMN} (see Appendix)
We replace the integration over $J-1$ angular variables $\theta_k$ by
a constant factor $I_{J-1}(2\pi)$ and fixed even-interval points $\theta_k=2\pi k/J$
and we have
\begin{equation}
{\cal A}
=\frac{(2\pi)^2}{J^2}
e^{-im2\pi p/J}e^{in2\pi q/J}{\cal A}_0.
\end{equation}
This amplitude implies that the existence of a coupling
between the closed string state (\ref{11excitation}) and the single trace operator
\begin{equation}
\sum_{p,q=1}^Je^{-2\pi i(mp/J-nq/J)}
\tr({\cal Z}^{J-p}\phi_i{\cal Z}^{p-q}\phi_j{\cal Z}^q).
\end{equation}
Of cause, when $m\neq n$, this amplitude vanishes because of the summation of $p$ and $q$.
Generalization to the case with arbitrary number of excitations
are straightforward.
This result coincide with the suggestion in \cite{BMN}.

In the argument above, we have assumed $m,n\geq1$.
It is also possible to consider `level 0' excitation
by adding a factor $X^i$ to the vertex operator.
The contraction between this factor and a vertex on the boundary is
\begin{equation}
\langle X^i(0)\partial_\perp X^i(e^{i\theta})\rangle=1.
\end{equation}
Thus, this corresponds to an insertion of the scalar field $\phi_i$
in the single trace operator without phase factor like
\begin{equation}
\sum_{p=1}^J\tr({\cal Z}^{J-p}\phi_i{\cal Z}^p).
\label{Zphi}
\end{equation}
The factor $X^i$ in the closed string vertex
changes the wave function $f_J$ rather than the oscillator parts $V\wt V$.
This is consistent with the fact that the operator (\ref{Zphi}) is a chiral primary
and is related to a Kaluza-Klein mode of supergravity fields on ${\bf S}^5$
via the AdS/CFT correspondence.

Inclusion of excitations by oscillators of Neumann directions
is straightforward.
The Green function of a scalar field on the unit disk with Neumann boundary condition is
\begin{equation}
\langle\phi(w)\phi(z)\rangle
=-\log|w-z|-\log|1-wz^\ast|+\frac{1}{2}(|w|^2+|z|^2).
\end{equation}
From this, we can obtain correlation functions between
operator $\partial^nX^\mu$ or $\wt\partial^nX^\mu$ at the center of
the disk and vertex operator $\partial_\parallel X^\mu$ on the
boundary as
\begin{equation}
\frac{1}{n!}\langle\ \partial^nX^\mu(0)\ \partial_\parallel X^\nu(e^{i\theta})\ \rangle
=-i\eta^{\mu\nu}e^{-in\theta},\quad
\frac{1}{n!}\langle\ \wt\partial^nX^\mu(0)\ \partial_\parallel X^\nu(e^{i\theta})\ \rangle
=i\eta^{\mu\nu}e^{in\theta}.
\end{equation}
Therefore, an excitation $\partial^n X^\mu$ corresponds to
an insertion of the gauge field $A_\mu$ with a phase factor like
\begin{equation}
\sum_{p=1}^J \tr(\cdots A_\mu\cdots)e^{-2\pi inp/J},
\label{Ainsertion}
\end{equation}
where $p$ is a position of $A_\mu$ inserted.
A right moving excitation $\wt\partial^nX^\mu$ corresponds
to a similar insertion with an opposite phase.
The operator (\ref{Ainsertion}) is gauge invariant if $n\geq1$.
Indeed we can rewrite (\ref{Ainsertion}) up to constant factor as
\begin{equation}
\sum_{p=1}^J \tr(\cdots[A_\mu,\cdot]\cdots)e^{-2\pi inp/J}.
\label{commAZ}
\end{equation}
Because we assume vanishing open string momenta, the commutator $[A_\mu,{\cal O}]$ is
equivalent to the covariant derivative $D_\mu{\cal O}=\partial_\mu{\cal O}+[A_\mu,{\cal O}]$
and is gauge covariant.
Therefore, the trace (\ref{commAZ}) is a gauge invariant operator.

If $n=0$, however, the expression (\ref{commAZ}) identically vanishes.
We can also see this on the CFT side as follows:
Let us add a level zero excitation factor $X^\mu$ to a closed string vertex
and compute the amplitude.
Unlike the case of Dirichlet boundary condition,
it has vanishing contraction with a vertex on the boundary.
\begin{equation}
\langle\ X^\mu(0)\ \partial_\parallel X^\nu(e^{i\theta})\ \rangle=0.
\end{equation}
Therefore, we obtain a vanishing coupling between the closed string and 
the operator with gauge field insertion without phase factor.
\section{Fermionic excitations}
Let us discuss fermion excitations of closed strings.
In the previous section, we have not specified the vacuum state (\ref{simplest}).
Let us begin with determining it.
Among $256$ ground states,
only one state gives non-vanishing one point function $\langle{\cal F}\rangle$.
To determine such a ${\cal F}$, we can use symmetries of the system.
The rotational symmetry transverse to the spacelike light-cone is
$SO(1,3)\times SO(4)$.
By taking T-duality along $X^4$, $X^5$, $X^6$ and $X^7$, this symmetry is enhanced into $SO(1,7)$.
Let us refer to this symmetry as $\wt{SO(1,7)}$.
In this section, we use this T-dual picture to make the $\wt{SO(1,7)}$ symmetry manifest.
To have non-vanishing $\langle{\cal F}\rangle$,
we should take an $\wt{SO(1,7)}$ singlet as ${\cal F}$.
There are two singlet physical states.
One is an NS-NS state (${\cal F}=e^{-\phi-\wt\phi}\sum_{\mu=0}^7\psi_\mu\wt\psi^\mu$)
and the other is an R-R state (${\cal F}=e^{-\phi/2-3\wt\phi/2}SC\Gamma^z\wt S$).
However, one point function of the R-R singlet state vanishes because it carries angular momentum $J=1$.
Therefore, the `vacuum state' should be the NS-NS singlet state.
In the original picture before taking the T-duality, this operator is represented as
\begin{equation}
{\cal F}=e^{-\phi-\wt\phi}\left(\sum_{\mu=0}^3\psi_\mu\wt\psi^\mu-\sum_{i=4}^7\psi^i\wt\psi^i\right).
\end{equation}
This state should be identified with the unique ground state of a closed string
in the PP-wave background\cite{Metsaev}.
In the GS formalism, all the other massless states are generated from it
by acting the zero modes of the GS fermions $\psi^a$ in $\bf 8_s$ representation of $\wt{SO(1,7)}$.
They belong to $\bf(8_v+8_c)\times(8_v+8_c)$ of $\wt{SO(1,7)}$
and are decomposed into irreducible representations as follows.
\begin{equation}
\begin{array}{cccc}
\mbox{NS-NS} & {\bf1+28+35_v}, &
\mbox{NS-R} & {\bf8_s+56_s}, \\
\mbox{R-R} & {\bf1+28+35_c}, &
\mbox{R-NS} & {\bf8_s+56_s}.
\end{array}
\end{equation}
This completely coincides with a set of the
antisymmetric products of from zero to eight $\bf8_s$'s.
Therefore, we guess the corresponding operators
in ${\cal N}=4$ SYM are ones with gaugino insertions without phase factor.
As the simplest case, let us consider one gaugino insertion.
The relevant closed string state must belongs to the $\bf8_s$ representation of
$\wt{SO(1,7)}$.
The vertex operator of the $\bf 8_s$ state in the R-NS sector is
\begin{equation}
{\cal O}_c=e^{-\phi/2-\wt\phi}\ol\zeta_{\dot a}(\Gamma_\mu)^{\dot a}{}_bV^b_{-1/2}\wt V_{-1}^\mu,
\label{8svertex}
\end{equation}
where the physical state condition for the polarization spinor $\ol\zeta_{\dot a}$ is
$\ol\zeta\Gamma^z=0$.
The amplitude we would like to compute is
\begin{equation}
\langle\ {\cal O}_c
\ \partial_\perp Z^\ast
\cdots
\ \partial_\perp Z^\ast
\ V_{-1/2}[\lambda_a]
\ \partial_\perp Z^\ast
\cdots
\ \partial_\perp Z^\ast
\ \rangle,
\end{equation}
where the gaugino vertex $V_{-1/2}[\lambda_a]$ is inserted into the $p$-th interval $[\theta_p,\theta_{p+1}]$.
Because the gaugino has no momentum,
any conditions are not imposed on $\lambda_a$.
\begin{figure}[htb]
\centerline{\epsfbox{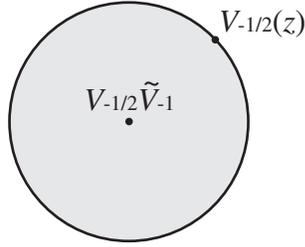}}
\caption{A worldsheet with one fermion vertex insertion.
All the insertions of bosonic operators on the boundary are omitted.}
\label{fermi0.eps}
\end{figure}
The amplitude is decomposed into several sectors.
The sectors of boson fields $X^I$ and the conformal ghost
are the same with what is discussed
in the last section.
In addition to it, we now have
the superconformal ghost sector
\begin{equation}
\langle\ e^{-\phi(0)/2}
\ e^{-\wt\phi(0)}
\ e^{-\phi(z)/2}
\ \rangle\propto\frac{1}{z^{1/4}},
\end{equation}
and the matter fermion sector
\begin{equation}
\langle\ \ol\zeta_{\dot a}(\Gamma^i)^{\dot a}{}_bS^b(0)
\ \wt\psi^i(0)
\ \lambda_aS^a(z)
\ \rangle\propto\frac{1}{z^{3/4}}(\ol\zeta C\lambda),
\end{equation}
where $C$ is the charge conjugation matrix in ten-dimension.
Because the closed string polarization spinor $\ol\zeta$ carries $J=-1/2$,
only half of $\lambda$ with $J=1/2$ participate in this process.
These factors are combined into
\begin{equation}
{\cal A}\propto\int d^{J-1}\theta\int_{[p]}dz\frac{1}{z}(\ol\zeta C\lambda),
\end{equation}
where $[p]$ is the arc between two points $\theta_{p+1}$ and $\theta_p$ on the boundary.
By means of the dilute gas approximation, we obtain the following $p$-independent result.
\begin{equation}
{\cal A}\propto\ol\zeta C\lambda.
\end{equation}
This implies that the closed string state (\ref{8svertex}) couples to the gauge invariant operator
\begin{equation}
\sum_{p=1}^J\tr({\cal Z}^{J-p}\lambda^{J=1/2}{\cal Z}^p),
\label{onelambda}
\end{equation}
where $\lambda^{J=1/2}$ represent eight components satisfying $\lambda\Gamma^{z^\ast}=0$.
Because the NS-R closed string vertex
\begin{equation}
{\cal O}_c=e^{-\phi/2-\wt\phi}\ol\zeta_{\dot a}(\Gamma_\mu)^{\dot a}{}_bV_{-1}^\mu\wt V^b_{-1/2},
\label{theother}
\end{equation}
also couples to the operator (\ref{onelambda}),
a closed string state coupled to the boundary operator (\ref{onelambda}) is
a linear combination of the R-NS state (\ref{8svertex})
and the NS-R state (\ref{theother}).
The other combination of these two states would couples to
a boundary operator with seven fermion insertions, which also belongs to the $\bf8_s$ representation
of $\wt{SO(1,7)}$.
In this way, we can establish a relation between the $256$ physical massless states
of a closed string and operators with gaugino insertions without phase factors.

Next, let us discuss excitations by fermion non-zero modes.
In NSR formalism,
the excitation is realized by a contour integral of a fermion vertex.
The level $n$ fermion excitation of a vertex ${\cal O}_c$ is given by
\begin{equation}
{\cal O}'_c(0)=\oint dw\frac{1}{w^n}V_{1/2}[u^a](w) {\cal O}_c(0).
\label{contour}
\end{equation}
We here neglect the extra terms necessary for the vertex to be BRS invariant
as we have done for bosonic excitations.
The vertex operator $V_{1/2}[u_a]$ in picture $1/2$ is defined by
\begin{equation}
V_{1/2}[u_a]=e^{-\phi/2}u_a(\Gamma^z)^a{}_{\dot b}\ol S^{\dot b}\partial Z^\ast+\cdots,
\end{equation}
where `$\cdots$' means terms irrelevant to our calculation.
When ${\cal O}_c$ is the ground state vertex (\ref{simplest}) and $n=0$,
${\cal O}_c'$ is fermion vertex given in (\ref{8svertex}).
As a simplest case, let us consider one fermion insertion
given by (\ref{contour}) with ${\cal O}_c$ in (\ref{simplest}).
In what follows, we omit all the bosonic open string operators because
they just give the factor we have discussed in the previous section.
The relevant part of the amplitude is
\begin{equation}
{\cal A}\propto\int d^{J-1}\theta\int_{[p]}dz\oint dw\frac{1}{w^n}
\eta_{\mu\nu}\langle\ V_{-1}^\mu(0)
\ V_{-1}^\nu(\infty)
\ V_{+1/2}[u_a](w)
\ V_{-1/2}[\lambda_a](z)
\ \rangle
\end{equation}
with the polarization $u$ satisfying
\begin{equation}
u\Gamma^{z^\ast}=0.
\end{equation}
Although we also need boson excitations to satisfy the level matching condition,
we omit them because they just give the factor we obtained in the last section.
\begin{figure}[htb]
\centerline{\epsfbox{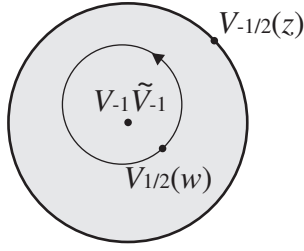}}
\caption{A world sheet with one fermion insertion on the boundary and
closed string vertex with a fermionic excitation.
All the insertions of bosonic operators on the boundary are omitted.}
\label{fermi.eps}
\end{figure}
The matter fermion sector is
\begin{equation}
\langle\ \psi_\mu(0)
\ \wt\psi^\mu(0)
\ u\Gamma^z\ol S(w)
\ \lambda S(z)
\ \rangle\propto
\frac{w+z}{w^{1/2}z^{1/2}(w-z)^{5/4}}(u\Gamma^zC\lambda).
\end{equation}
Because of the coupling $u\Gamma^zC\lambda$, only $J=1/2$ components of $\lambda$
satisfying $\lambda\Gamma^{z^\ast}=0$ participate in the interaction.
The superconformal ghost sector is
\begin{equation}
\langle e^{-\phi(0)}e^{-\wt\phi(0)}
e^{\phi(w)/2}e^{-\phi(z)/2}\rangle\propto
\frac{w^{1/2}(w-z)^{1/4}}{z^{1/2}}.
\end{equation}
Furthermore, we have a factor $1/w$ coming from
a contraction between $\partial Z^\ast$ in $V_{1/2}$ and $f_J(Z)$.
The correlation function is
obtained as a product of these factors and
boson factor discussed in the previous section.
By using the dilute gas approximation,
the amplitude is obtained as
\begin{equation}
{\cal A}\propto\int d^{J-1}d\theta\int_{[p]}dz\oint dw\frac{1}{w^n}\frac{z+w}{wz(z-w)}(u\Gamma^zC\lambda)
\propto e^{-2\pi inp/J}(u\Gamma^zC\lambda).
\end{equation}
Now we are omitting the factor from boson excitations.
The amplitude has a phase factor $e^{-2\pi inp/J}$ depending on the position of
the fermion insertion.
This implies that the closed string vertex ${\cal O}'_c$ couples to the
boundary operator
\begin{equation}
\sum_{p=1}^J\tr(\cdots \lambda^{J=1/2} \cdots)e^{-2\pi inp/J},
\end{equation}
where $p$ is the position of the $\lambda^{J=1/2}$ insertion.
This is exactly what suggested in \cite{BMN}.

\section{Conclusions}
In this paper we investigated large angular momentum closed strings
colliding with D3-branes and decaying into large number of open strings.
As couplings between closed strings
and gauge invariant operators on D3-branes,
the string states-boundary operators correspondence
suggested in \cite{BMN} is reproduced
completely for
states without fermionic excitation.
Concerning states including fermionic excitations,
we gave explicit computation of amplitudes
only for the case of one fermion excitation.
It is plausible that the correspondence for general states would be
reproduced in the same way.

We briefly comment on fields carrying $d-J=2$.
In processes consistent with the least number condition,
only $d-J=0$ (${\cal Z}$) and $d-J=1$ fields ($A_\mu$, $\phi_i$, $\lambda^{J=1/2}$)
participate in the interactions.
Although single trace operators including $d-J=2$ fields (${\cal Z}^\ast$, $\lambda^{J=-1/2}$)
also have non-vanishing couplings with closed strings,
the operators `decay' into other operators consisting of
less number of fields.\cite{BMN}
For example, let us consider the operator $\tr({\cal Z}^\ast{\cal Z}^{J+1})$.
It is expressed as
insertion of $J+1$ $\partial_\perp Z^\ast$ and one $\partial_\perp Z$ on the disk boundary.
Because the vertex $\partial_\perp Z$ can be contracted with one of $\partial_\perp Z^\ast$,
the operator cannot satisfy the least number condition.

After Ref.\cite{BMN},
many works has appeared discussing the pp-wave limit of
orbifold \cite{enhanced,GO,ZS,takatera,ASJ,FK,moose,quiver}
and open strings in the pp-wave background\cite{chuho,open,LeePark}
along the line of \cite{BMN}.
We hope that our CFT analysis is also useful to study these cases.

\section*{Acknowledgements}
We would like to thank T.~Takayanagi for
a nice lecture about recent development of
the pp-wave/gauge theory correspondence.

\appendix
\section{Dilute gas approximation}
Consider the following integral:
\begin{eqnarray}
A=\int_0^xd^nx\int_{x_k}^{x_{k+1}}dyf(y),
\end{eqnarray}
where $\int_0^xd^nx$ is defined by
\begin{eqnarray}
\int_0^xd^nx=\int_0^xdx_n\int_0^{x_n}dx_{n-1}\cdots\int_0^{x_2}dx_1.
\end{eqnarray}
The function $f(y)$ depends only one variable $y$.
Integration over variables $x_i$ gives
\begin{equation}
A=\int_0^xI_k(y)I_{n-k}(x-y)f(y)dy=\int_0^x\rho(y)f(y)dy.
\label{rhofint}
\end{equation}
where the density function $\rho(y)$ is defined by
\begin{equation}
\rho(y)
=I_k(y)I_{n-k}(x-y)
=\frac{y^k(x-t)^{n-k}}{k!(n-k)!}.
\end{equation}
This function is expanded around its maximum point $y=y_0$ as
\begin{equation}
\rho(y_0+\epsilon)=\rho(y_0)\left(1-\frac{\epsilon^2}{2x^2}\frac{n^3}{k(n-k)}+{\cal O}(\epsilon^3)\right),\quad
y_0=\frac{k}{n}x.
\end{equation}
If $n$ is sufficiently large,
we can estimate the integral (\ref{rhofint}) by steepest dissent approximation and we obtain
\begin{equation}
A=\frac{x}{n}\sqrt{\frac{2\pi k(n-k)}{n}}\rho(y_0)f(y_0).
\label{Arho}
\end{equation}
On the other hand,
if both $k$ and $n-k$ are the same order of magnitude as $n$,
the maximum of the function $\rho(y)$ is estimated by
the Stirling formula
$z!/z^z=\sqrt{2\pi z}e^z(1+{\cal O}(z^{-1}))$.
\begin{equation}
\rho(y_0)=
\frac{k^k}{k!}
\frac{(n-k)^{n-k}}{(n-k)!}
\frac{n!}{n^n}
\frac{x^n}{n!}\nonumber\\
=\sqrt{\frac{n}{2\pi k(n-k)}}\frac{x^n}{n!}(1+{\cal O}(n^{-1})).
\label{maxrho}
\end{equation}
By substituting (\ref{maxrho}) into (\ref{Arho}),
we obtain a simple expression for $A$.
\begin{equation}
A=\frac{x}{n}f\left(\frac{kx}{n}\right)I_n(x)
\label{Aformula}
\end{equation}

With the help of this formula,
we can easily estimate the following integral of a function
depending on two variables.
\begin{equation}
B=\int_0^xd^nx\int_{x_k}^{x_{k+1}}dy\int_{x_l}^{x_{l+1}}dzf(y,z)
\end{equation}
We can carry out $x_i$ integration to obtain
\begin{eqnarray}
B=\int_0^xdy\int_0^ydzI_{n-k}(x-y)I_{k-l}(y-z)I_l(z)f(y,z).
\end{eqnarray}
By the formula (\ref{Aformula}) we can carry out $z$ integration and
the result is
\begin{equation}
B=I_{n-k}(x-y)I_k(y)\times\frac{y}{k}f\left(y,\frac{ly}{k}\right).
\end{equation}
We again use the formula (\ref{Aformula}) to integrate over $y$
and obtain
\begin{equation}
B=\frac{x^2}{n^2}f\left(\frac{kx}{n},\frac{lx}{n}\right)I_n(x)
\end{equation}

Generalization to the case of the integrated function depending on an arbitrary number of variables
is
\begin{equation}
\int_0^xd^nx\prod_{i=1}^p\int_{x_{k_i}}^{x_{k_i+1}}dy_i f(y_1,y_2,\ldots,y_p)
=\frac{x^p}{n^p}f\left(\frac{k_1x}{n},\frac{k_2x}{n},\ldots,\frac{k_px}{n}\right)I_n(x)
\end{equation}
This formula is expressed by the statement:
``we can replace the $x_i$ integrals with constant factor
$I_n(x)$ and fixed even-interval points $x_k=(k/n)x$.''


\end{document}